\documentclass[fleqn,usenatbib]{mnras}

\usepackage[T1]{fontenc}

\DeclareRobustCommand{\VAN}[3]{#2}
\let\VANthebibliography\thebibliography
\def\thebibliography{\DeclareRobustCommand{\VAN}[3]{##3}\VANthebibliography}

\usepackage{graphicx}	
\usepackage{amsmath}	
\usepackage{amssymb}	
\usepackage{pdflscape}  
\usepackage{enumitem}   

\PassOptionsToPackage{normalem}{ulem}
\usepackage{ulem}

\usepackage{newtxtext,newtxmath}

\title[Validated Beam Models for an EoR Detection]{The Necessity of Individually Validated Beam Models for an Interferometric Epoch of Reionization Detection}

\author[A. Chokshi et al.]{A. Chokshi,$^{1,2,3}$\thanks{E-mail:\href{mailto:achokshi@student.unimelb.edu.au}{achokshi@student.unimelb.edu.au}}
N. Barry,$^{2, 4}$
J. L. B. Line,$^{2,5}$
C. H. Jordan,$^{2,5}$
B. Pindor,$^{1,2}$
R. L. Webster$^{1,2}$
\\
$^{1}$The University of Melbourne, School of Physics, Parkville, VIC 3010, Australia\\
$^{2}$ARC Centre of Excellence for All Sky Astrophysics in 3 Dimensions (ASTRO 3D)\\
$^{3}$CSIRO Astronomy and Space Science (CASS), PO Box 76, Epping, NSW 1710, Australia\\
$^{4}$School of Physics, University of New South Wales, Sydney, NSW 2052, Australia\\
$^{5}$International Centre for Radio Astronomy Research, Curtin University, Perth, WA 6845, Australia }

\date{Accepted XXX. Received YYY; in original form ZZZ}

\pubyear{2015}

\begin{document}
\label{firstpage}
\pagerange{\pageref{firstpage}--\pageref{lastpage}}
\maketitle

\begin{abstract}
A first statistical detection of the 21-cm Epoch of Reionization (EoR) is on the horizon, as cosmological volumes of the Universe become accessible via the adoption of low-frequency interferometers.  We explore the impact which non-identical instrumental beam responses can have on the calibrated power spectrum and a future EoR detection. All-sky satellite measurements of Murchison Widefield Array (MWA) beams have revealed significant sidelobe deviations from cutting-edge electromagnetic simulations at the $\sim$10$\%$ zenith power level. By generating physically motivated deformed beam models, we emulate real measurements of the MWA which inherently encode the imprints of varied beams. We explore two calibration strategies: using a single beam model across the array, or using a full set of deformed beams. Our simulations demonstrate beam-induced leakage of foreground power into theoretically uncontaminated modes, at levels which exceed the expected cosmological signal by factors of over $\sim$1000 between the modes $k$=0.1-1 $h\mathrm{Mpc}^{-1}$. We also show that this foreground leakage can be mitigated by including measured models of varied beams into calibration frameworks, reducing the foreground leakage to a sub-dominant effect and potentially unveiling the EoR. Finally, we outline the future steps necessary to make this approach applicable to real measurements by radio interferometers.
\end{abstract}




\section{Introduction}
\label{sec:intro}

The past decade has seen the adoption of relatively simple, large interferometric arrays as powerful tools for the investigation of the low-frequency radio sky. These aperture arrays are generally constructed from sets of simple metal dipoles, coherently synthesised to achieve high angular resolution imaging over unprecedented wide fields-of-view. Such telescopes are often designed to have a large number of receiving elements (tiles or stations), each constructed from a number of identical dipoles, with theoretically identical sensitivities across the sky.

Low-frequency radio interferometers with large collecting areas can sample many modes on the sky, allowing them to search for faint, cosmic signals.  The Murchison Widefield Array (MWA, \citealt{Tingay_MWA_2013, Wayth_MWA_2018}), the Hydrogen Epoch of Reionization Array (HERA,  \citealt{deboer_hydrogen_2017}), the New Extension in Nançay Upgrading LOFAR (Nenufar, \citealt{zarka_low-frequency_2020, Munshi_nenuFAR_2024}), the Low Frequency Array (LOFAR, \citealt{van_haarlem_lofar:_2013}), and the Long Wavelength Array (LWA, \citealt{eastwood_21_2019}) are all searching for cosmic signals below 200\,MHz. 



Understanding the telescope's varied sensitivity across the sky, or primary beam response, is a crucial part of the inherent calibration process. Beam sensitivity measurements show that this often differs from the instrumental simulations, especially in attenuated parts of the beam. Sensitivity measurements have been made with the MWA \citep{bowman_field_2007,neben_measuring_2015,Line_ORBCOMM_2018,Chokshi_SAT_MAPS_2021}, with LOFAR \citep{ninni_comparison_2020}, and with HERA \citep{neben_hydrogen_2016, nunhokee_measuring_2020}.



Ideally, the beam shape of each interferometric station or tile is identical, enabling massive computational simplifications during beam calibration. 
However, the realities of dipole failure and other environmental perturbations breaks this assumption (e.g. as measured by \citealt{Line_ORBCOMM_2018,Chokshi_SAT_MAPS_2021}) and requires more complicated calibration schemes to be considered in the pursuit of high fidelity science. This may prove costly for extremely large arrays, especially future telescopes like the Square Kilometre Array (SKA-Low, \citealt{mellema_reionization_2013, koopmans_cosmic_2015}). 



\begin{figure}
    \includegraphics[width=\columnwidth]{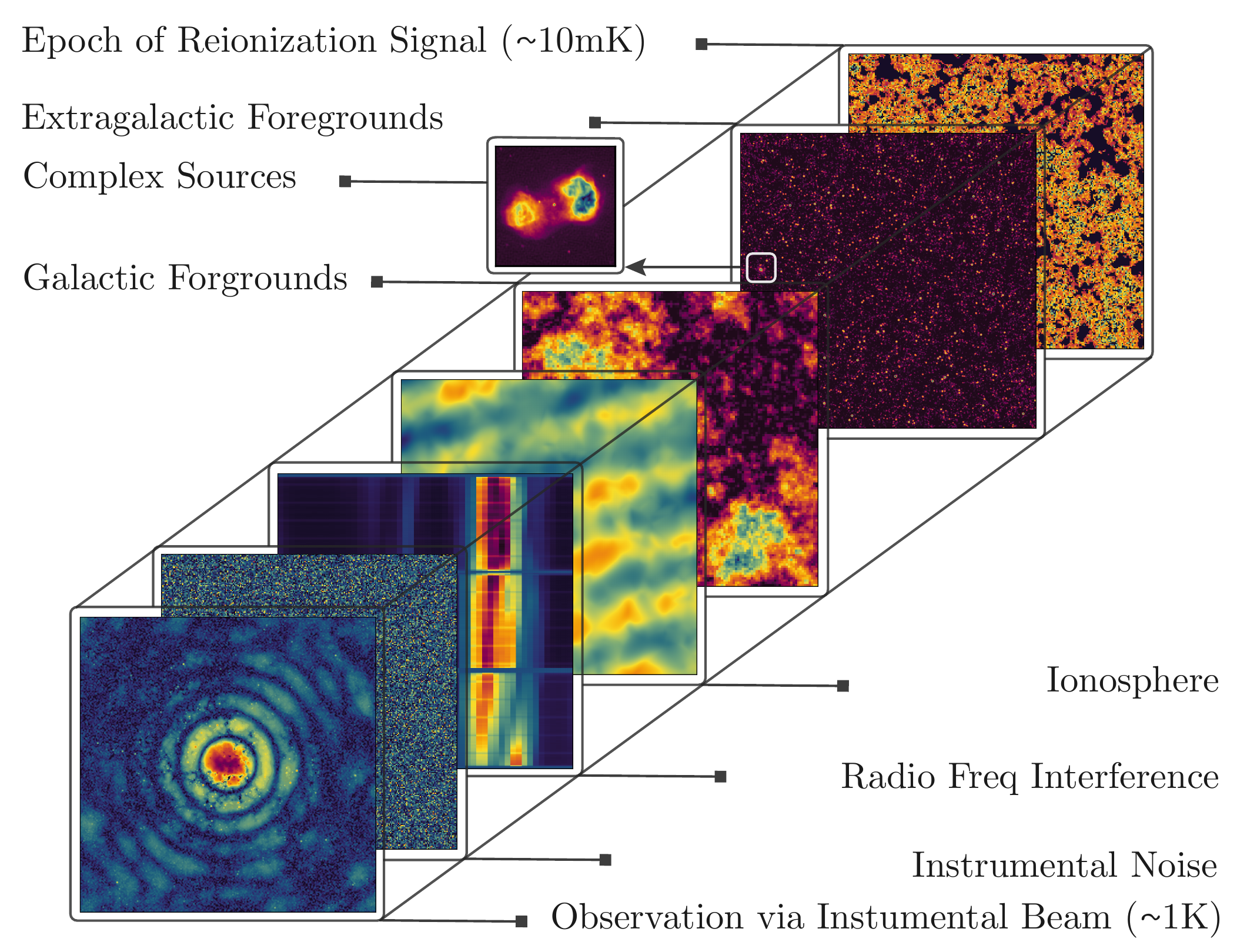}
    \caption{A schematic representation of the primary contributing components captured in an standard EoR observation (inspired by figures in \citealt{jelic_firegrounds_2008}), spanning five orders of magnitudes from the faint cosmological signal, to foreground, terrestrial and instrumental effects.}
    \label{fig:slices}
\end{figure}

The precision of calibration is particularly crucial for power spectrum measurements of the 21-cm Epoch of Reionisation (EoR) signal. This cosmological, redshifted signal is expected to be up to five orders of magnitude fainter than the various  foregrounds (see Figure \ref{fig:slices}) \citep[e.g.][]{oh_peng_foregrounds_2003, santos_eor_2005, pober_wedge_2013, yatawatta_lofar_eor_2013}, but will naturally separate in Fourier space due to its varying spectral structure. However, calibration can impart varying structure on otherwise spectrally smooth foregrounds, clouding the EoR measurement (e.g. \citealt{barry_calibration_2016, patil_systematic_2016,byrne_fundamental_2019}). 

Calibration errors in the context of beam variations have been explored within simulation. Redundant calibration, where tile parameters are reduced from multiple measurements of the same mode, is particularly susceptible to variations in antennas and their placement \citep{joseph_bias_2018,orosz_mitigating_2019,choudhuri_patterns_2021,kim_impact_2022}. Sky calibration, where tile parameters are reduced from comparisons between measurements and full-sky models, is also affected by unaccounted broken dipoles within tiles or stations \citep{joseph_calibration_2020}. Given the computational complexity of unique beams in analyses, these studies explore discrete variations. 

We show a more complete picture of the effects of beam variation within sky calibration of MWA Phase II, using actual beam measurements to inform our simulations. We have 14 dual polarisation measurements of true beam variation from \citet{Chokshi_SAT_MAPS_2021}, and we produce simulations which use these measurements to modify the dipole gains within a tile on a floating-point level to match. While our simulations are still encoding discrete variation representative of 14 measurements, it adds to work that was previously binary in nature \citep{joseph_calibration_2020}. This gives a more realistic portrayal of expected errors from an instrument that has been in the field for over a decade. Given our evidence-based beam variations and our analysis framework, deformed beams may be the cause of current limiting systematics in recent MWA limits \citep{trott_deep_2020,rahimi_epoch_2021}. 

In Section~\ref{sec:sat_maps}, we describe how we build optimal beam maps via satellite measurements from \citet{Chokshi_SAT_MAPS_2021} for 14 tiles. In Section~\ref{sec:cali}, we take these optimal maps and forward model them through a simulation and calibration framework  which is representative of real data analysis. We summarise our power spectrum metric in Section~\ref{sec:ps} and investigate the effects of performing calibration with and without knowledge of the deformed beams in Section~\ref{sec:results} and compare the results in power spectrum space. We summarise our conclusions in Section~\ref{sec:conclusions}.



\section{Optimal Satellite Beam Maps}
\label{sec:sat_maps}

The Fully Embedded Element (FEE) beam model \citep{Sutinjo_FEE_2015, Sokolowski_FEE_2017} is a cutting-edge numerical electro-magentic simulation of the MWA tile response using FEKO\footnote{\url{http://www.feko.info}}. The FEE beam model incorporates a number of significant improvements over the previous analytic representations of the beam, including mutual coupling between the the multiple dipoles in the tile and a model of the electromagnetic effects of the soil below the tile. 

The FEE simulations represent a tile under ideal conditions. Unfortunately, the arid conditions at the MWA site, and its remote location lead to a range of environmental factors which perturb beam models away from the FEE standard. In-situ, all-sky measurements of MWA beam shapes using communication and weather satellites have shown that the measured beam shapes differ from the FEE model, particularly away from zenith and within the sidelobes of the beams, at a $\sim$10$\%$ level \citep[see,][]{Line_ORBCOMM_2018, Chokshi_SAT_MAPS_2021}. The dual polarised beam shapes of 14 MWA tiles were measured by \cite{Chokshi_SAT_MAPS_2021}, creating all-sky HEALPix \citep{Gorski_healpix_2005} maps with a angular resolution of 110 arcminutes at 137 MHz. These maps were created by an open-source \texttt{Python} package called \texttt{EMBERS} \citep{Chokshi_EMBERS_2020}, and are available online. The direct incorporation of these measured beam maps into standard calibration software is hindered by their low resolution and narrow frequency bandwidth. 

The FEE beam model has 16 variable dipole amplitude parameters per polarisation, which can each be tuned to weight the contribution of dipoles to the MWA tile. Typically, all dipole amplitudes are set to one, representing a perfect tile, with the occasional tile having a single dipole set to zero indicating the presence of a malfunctioning or flagged dipole (occurring in $\sim$20--40$\%$ of all tiles at any given time, see \citealt{joseph_calibration_2020}). This predominantly occurs due to the failure of the primary low noise amplifier (LNA) within the central column of the MWA dipoles as they gradually degrade upon contact with the slightly acidic local soil. 

Our proposed method of incorporating more complex and perturbed beam models is to use the measured satellite beam maps to determine the optimal set of 16 dipole amplitudes, which best reproduce the measurements. This does not address the issue of extrapolating the narrow bandwidth satellite measurements at 137~MHz, as most Epoch of Reionization searches are conducted across the 167-198 MHz band where Galactic \& extragalactic foregrounds and ionospheric effects are least dominant. Given the response of the MWA FEE beam, to first order, the scaling of these dipole amplitudes across frequency is considered to be linear. A study of the frequency scaling of these dipole parameters is beyond the scope of this work as it would likely involve drone measurements of the MWA beam patterns across the entire frequency band. In this work, we assume that it is valid to linearly extrapolate the dipole parameters recovered from 137 MHz satellite beam maps across the 167-198 MHz band where EoR observations are conducted. 

The beam maps from \cite{Chokshi_SAT_MAPS_2021} are available\footnote{\url{https://github.com/amanchokshi/MWA-Satellite-Beam-Maps}} in the form of HEALPix maps of two types. The first represents a median satellite map, with pixel values averaged over all satellite passes, while the second are error maps with pixel values representing the median absolute deviation (MAD) of all satellite passes. 

We define a likelihood function $\mathscr{L}$ which quantifies how similar the FEE model with 16 dipole amplitudes ($d_{0}:d_{15}$) is to the measured beam maps. The set of dipole parameters which correspond to the maximum likelihood estimator $\mathscr{L}_{\mathrm{max}}$ leads an optimised FEE model.

\begin{equation} \label{eq:likelihood}
    \mathscr{L} = -1 \cdot \ln \left\{ \sum_{i=1}^{N} \frac{\Big|\mathbb{S}_{i} - \mathbb{F}_{i}|_{d_{0}:d_{15}}\Big|^2}{\mu_{\mathbb{S}\,i}} \right\},
\end{equation}

\noindent where $\mathbb{F}$ is the FEE beam model evaluated on a HEALPix grid, with a set of 16 dipole amplitudes, using the GPU accelerated \texttt{hyperbeam}\footnote{\url{https://github.com/MWATelescope/mwa_hyperbeam}} package. $\mathbb{S}$ is the satellite beam map with $\mu_{\mathbb{S}\,i}$ being the MAD error map and $i$ the pixel indices. Pixels with FEE power lower than $-30$dB from zenith are masked out due to low signal to noise, along with the central 20$^\circ$ where bright satellites saturated the amplifiers used in \cite{Chokshi_SAT_MAPS_2021}, leading to a low confidence central region.

We use the Bayesian Information Criteria (BIC) as the metric for our optimal model selection, as it accounts for the number of free parameters and amount of data used in the model evaluation, where BIC is defined as:

\begin{equation} \label{eq:bic}
    \mathrm{BIC} = k\cdot \ln (n) - 2 \cdot \mathscr{L}_{\mathrm{max}},
\end{equation}

\noindent where \textit{k} is the number of free parameters in the model (16 in the case of the FEE beam model), \textit{n} is the number of data points used (number of unmasked HEALPix pixels in satellite beam maps) and $\mathscr{L}_{\mathrm{max}}$ being the maximum likelihood estimator. The model with the lowest BIC value corresponds to a set of 16 dipole amplitude parameters which best optimise the FEE model to the given satellite beam map.

\begin{figure}
    \includegraphics[width=\columnwidth]{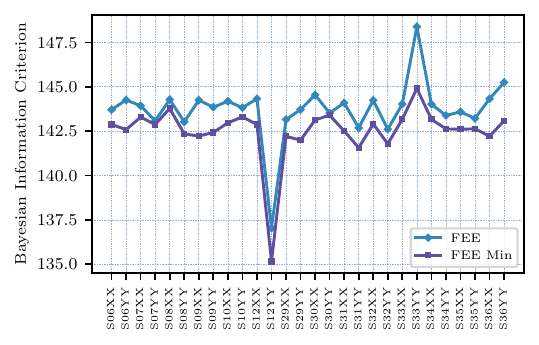}
    \caption{The best (lowest) BIC values obtained by the optimisation of the 16 dipole amplitude parameters in Eqn. \ref{eq:bic}, for the 14 dual polarised (XX, YY) MWA satellite beam maps available. The blue line (FEE), show the BIC value of the satellite map compared to the full FEE model, while the purple line (FEE Min) shows the BIC values for the optimised set of dipole amplitudes. Tile ``S12YY'' has lower BIC values due to sparse satellite coverage which led to a lower n in Eqn. \ref{eq:bic}.}
    \label{fig:beam_min_bic}
\end{figure}

\begin{figure*}
    \includegraphics[width=\textwidth]{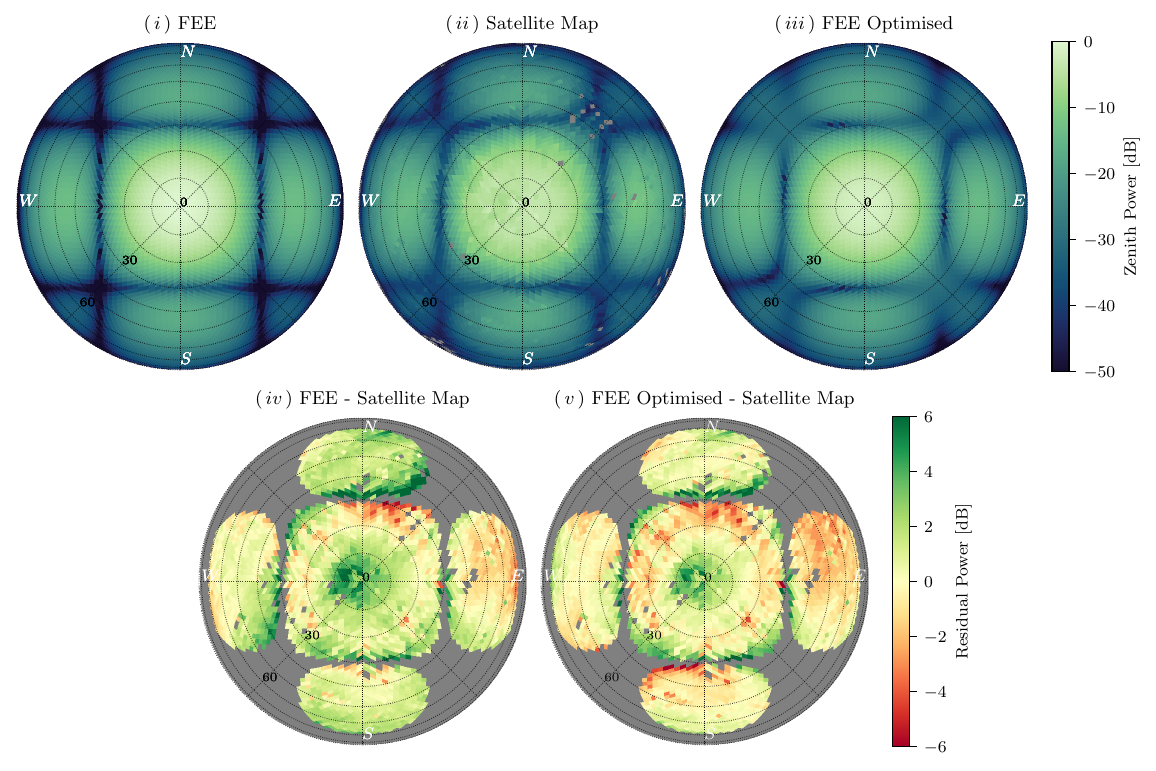}
    \caption{A study of the efficacy of the beam minimization procedure described in Section \ref{sec:sat_maps}, tested on MWA tile ``S06YY''. The top row (\textit{i, ii, iii}) represent the perfect FEE model, the measured satellite model, and the optimised FEE model using dipole amplitude parameters recovered by minimisation. The second row (\textit{iv, v}) depicts the residual power between the FEE, optimised FEE models and the satellite beam map. Panels (\textit{iii, v}) show that the optimised FEE model better matches the satellite beam maps (\textit{ii}), accurately capturing first-order beam deformations present in the satellite data.}
    \label{fig:S06YY}
\end{figure*}

\begin{figure*}
    \includegraphics[width=\textwidth]{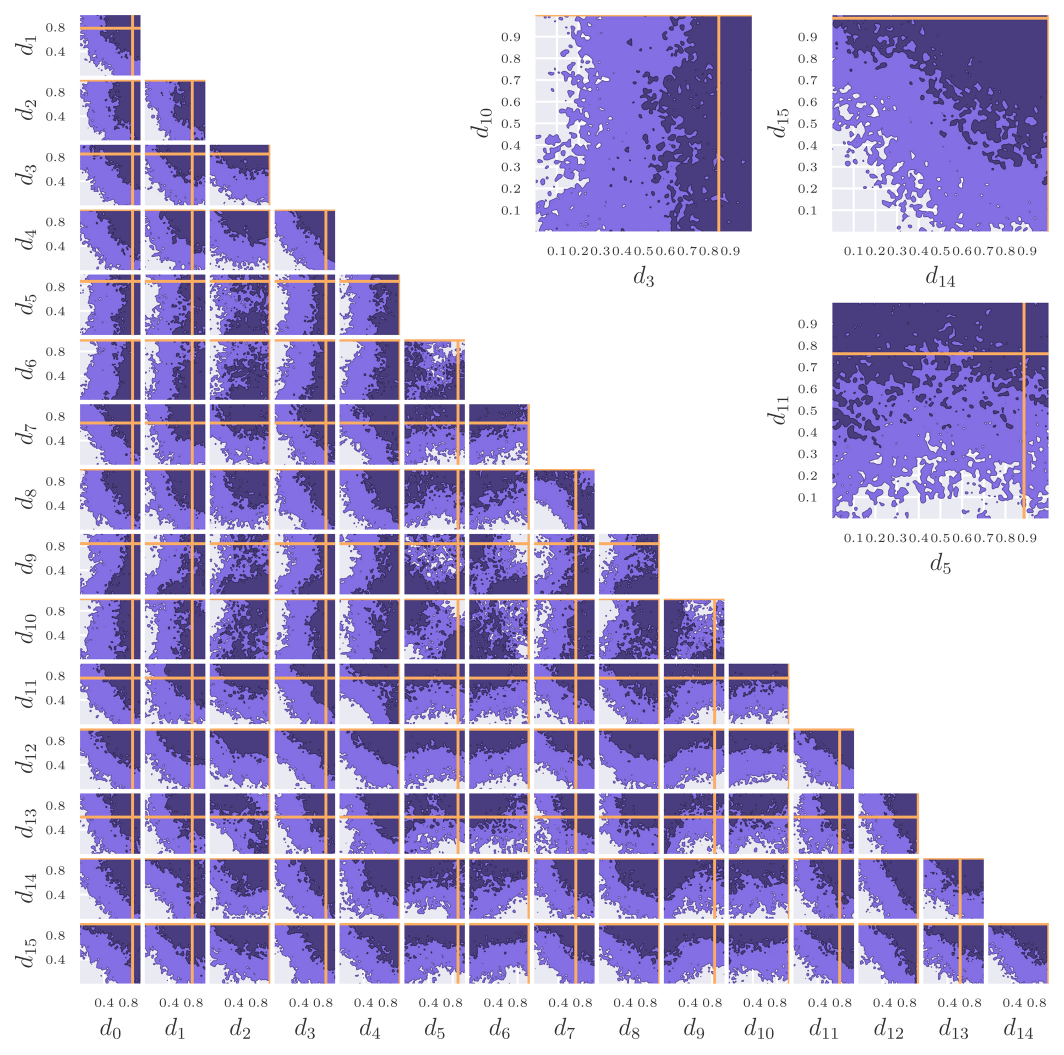}
    \caption{A MCMC analysis of MWA tile ``S06YY'' where purple contours represent 86\% and 39\% confidence levels respectively. The orange lines depict the results of beam minimisation from Section \ref{sec:sat_maps}. The insets on the top right focus on three sets of dipole pairs which display varying levels of degeneracy between parameter constraints.}
    \label{fig:mcmc}
\end{figure*}

Figure \ref{fig:beam_min_bic} shows the best BIC values obtained via the maximum likelihood estimator of Eqn. \ref{eq:bic} of an optimised FEE model (FEE Min - purple line), compared to the BIC value corresponding to a perfect FEE model (blue line), with all 16 dipole amplitudes set to 1. Figure \ref{fig:beam_min_bic} shows that the optimised FEE model is consistently preferred over the nominal FEE model, with improvements in BIC values of $\sim2$ across the board. In Figure \ref{fig:S06YY}, the 16 optimal dipole amplitudes for MWA tile ``S06'' in the North-South polarisation (henceforth ``S06YY''), recovered via the beam minimisation described above, are applied to the FEE model to quantify how well this process can reproduce measured MWA beam shapes. The top row ($i, ii, iii$) shows the perfect FEE beam model, the measured satellite beam model for tile ``S06YY'' and an optimised FEE model perturbed to best match the satellite map. Notice how the optimised FEE model ($iii$) has primary beam nulls which are less deep and distinct than the corresponding perfect FEE model, closely matching the satellite map ($ii$).  The bottom row ($iv, v$) depicts residuals between the FEE or optimised FEE model and the satellite beam map respectively, with regions of the FEE model lower than 30dB below zenith power being masked out due to low signal to noise. The residuals with the optimised FEE model ($v$) have visibly reduced gradients across the beam sidelobes, and better match the satellite map at the zenith.

An in-depth investigation into the distribution of optimal parameters in the 16-dimensional dipole amplitude space was performed for MWA tile ``S06YY'', using a Markov chain Monte Carlo (MCMC) method, with the likelihood defined in Eqn. \ref{eq:likelihood} and uniform, uninformative priors. The MCMC analysis was performed using a \texttt{Python} package called \texttt{EMCEE} \citep{Foreman_Mackey_EMCEE_2013}, with corner plots made using \texttt{ChainConsumer} \citep{Hinton_chain_consumer_2016}. Figure \ref{fig:mcmc} shows the result of the MCMC analysis, marginalised over pairs of parameters, with the purple contours representing 86\% (dark purple) and 39\% (light purple) confidence levels respectively. The orange lines represent the results of the beam minimisation described above, and shown in Fig. \ref{fig:S06YY}. While the results of the beam minimisation do concur with the central confidence contours in Fig. \ref{fig:mcmc}, large degeneracies are observed in certain pairs of parameters, representing a lack of tight constraints on some dipole amplitudes. The insets in the top right corner of Fig. \ref{fig:S06YY} show that for dipoles $d_3 \,\&\, d_{10}$, any possible value of $d_{10}$ is as valid. Similarly, for the dipole pair $d_5 \,\&\, d_{11}$, any value of $d_{11}$ is equally valid. In essence, this indicates that dipole $d_3 \,\&\, d_5$ place almost no constraints on dipoles $d_{10} \,\&\, d_{11}$, respectively. In contrast, the dipoles $d_{14} \,\&\, d_{15}$ constrain each other well, leading to much lower degeneracy between these parameters.

We observe that the pairs of dipole parameters which are often least well constrained include one of the four central dipoles. The FEE beam is used in a ``zenith normalised'' form, where zenith power is scaled to 1, with everything else being correspondingly scaled. We posit that the observed degeneracy in dipole amplitudes which arises from the central dipoles can be explained by the fact that variation in the central dipole amplitudes tend to scale the overall power without significant deviations in beam shape. The effect is mostly eliminated by the zenith normalization of the beam. In contrast, the 12 dipoles on the edge of a MWA tile have a more significant effect on beam shapes, leading to significant distortions in the beam sidelobes. The $\chi^2$ metric used in the beam minimisation and the MCMC analysis is only sensitive to global changes in the shape of the beam. The above procedure thus preferentially places most constraints on dipoles which affect the beam shape adversely. 

\section{Simulation \& Calibration Framework}
\label{sec:cali}



\subsection{Calibration \& Beams}
\label{ssec:cal_beams}

Each unique pair of antennas in an interferometer, separated by baseline $\mathbf{u}$, samples the sky brightness distribution $I(\mathbf{l}, \nu)$ by measuring of the complex visibility 

\begin{equation}
\begin{aligned}
V(\mathbf{u}, \nu) = \int g_{p} g^{*}_{q}\, b_{p}(\mathbf{l},\nu) b^{*}_{q} (\mathbf{l},\nu)\, I(\mathbf{l}, \nu)\, e^{-2\pi i\mathbf{u}\cdot \mathbf{l}}\, d^2\mathbf{l},
\end{aligned}
\label{eq: measurement_equation}
\end{equation}

\noindent where $\mathbf{l}$ is the sky coordinate vector, $\nu$ is the observing frequency, $g_{p}$ and $b_{p}$ are the complex-valued gain and voltage beam pattern of antenna $p$, respectively. Calibration of measured visibilities enables the accurate reconstruction of the true sky brightness distribution $I(\mathbf{l}, \nu)$. Equation \ref{eq: measurement_equation} demonstrates how each measured visibility $V(\mathbf{u}, \nu)$ contains the fundamental imprint of the constituent pair of receiving element beams. Traditional sky-based calibration \citep[e.g.,][]{Mitchell_RTS_2008, Salvini_Wijnholds_Cal_2014} minimises the squared differences between a measured visibility $V_{pq}^{\mathrm{data}}$ and a model visibility $V_{pq}^{\mathrm{model}}$ simulated from sky and beam models, to solve for unknown antenna complex-valued gains $g_p$ and $g_q$

\begin{equation}
\begin{aligned}
\chi^2  = \sum_{pq} \lvert V^{\mathrm{data}}_{pq} - g_{p} g_{q}^* V^{\mathrm{model}}_{pq} \rvert^2.
\end{aligned}
\label{eq: sky_cal}
\end{equation}

 This work investigates the effects of an imperfect representation of the instrumental beams during this critical calibration stage. 

\subsection{Fiducial Simulation}
\label{ssec: fid_sim}

To simulate a MWA array of 128 deformed tiles, 16 gain values are required per tile and polarization. \cite{Chokshi_SAT_MAPS_2021} measured all-sky dual-polarised beam maps of 14 fully polarized MWA tiles, and in Section \ref{sec:sat_maps} we determined the optimal gain parameters for each of their dipoles. For each dipole in a simulated deformed tile we make a random selection from the relevant 14 available gain parameters. This ensures that each of the 128 tiles has a physically motivated distortion model. This simulation framework can be used to emulate measurements made an interferometric array composed of deformed beams.

\begin{figure}
    \includegraphics[width=\columnwidth]{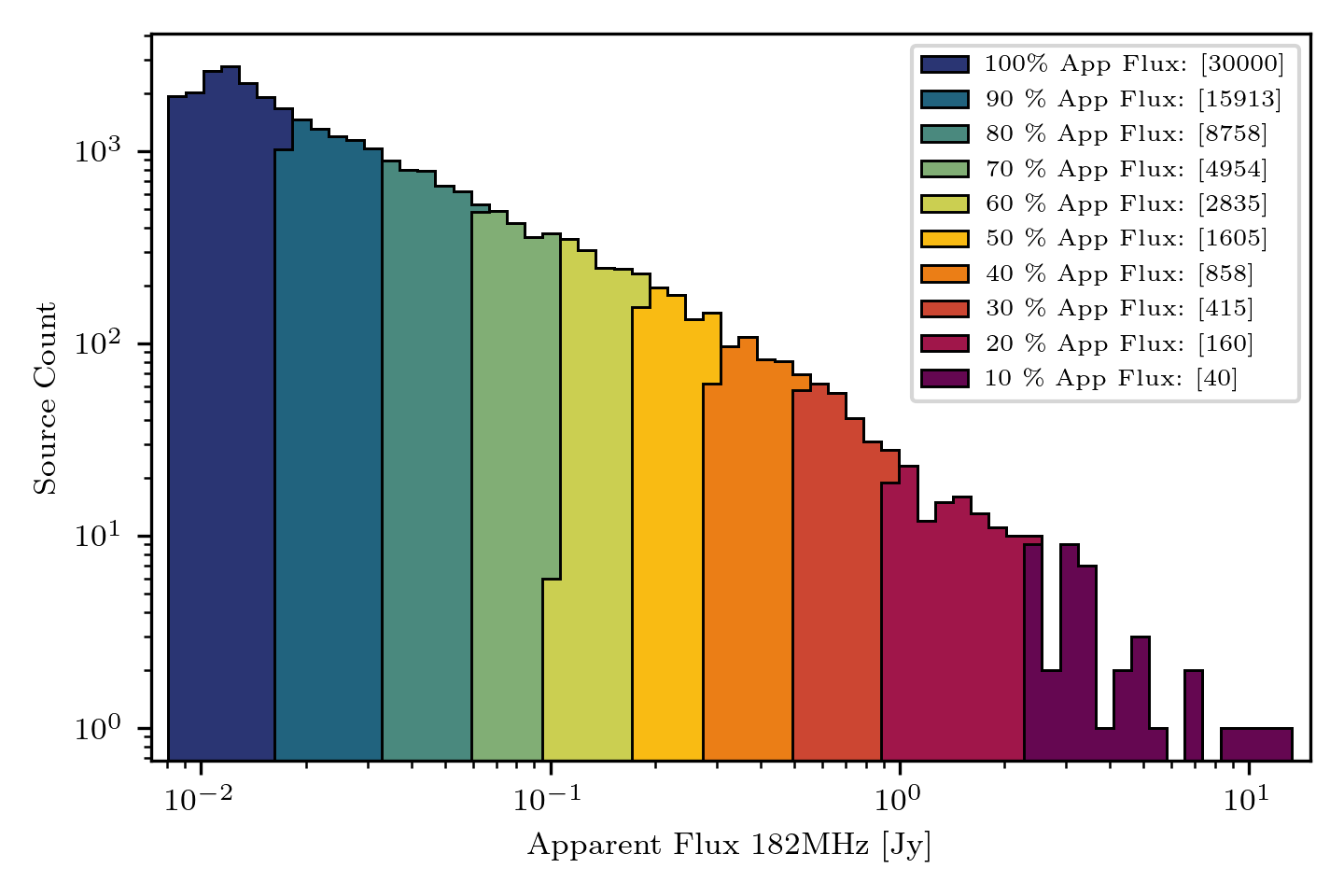}
    \caption{A histogram apparent brightness of all 30,000 sources included in this work, at 182MHz. Each coloured section represent 10\% of the integrated flux, from brightest on the right, to faintest on the left. }
    \label{fig:srclist}
\end{figure}

\texttt{hyperdrive}\footnote{\url{https://github.com/MWATelescope/mwa_hyperdrive}} (Jordan et al., in prep) is a cutting-edge sky-based calibration and simulation tool designed for the MWA, developed to be the successor to the Real Time System (RTS; \citealt{Mitchell_RTS_2008}). \texttt{hyperdrive} is used to create a noiseless simulation of the 30,000 brightest foreground sources (see Figure \ref{fig:srclist}) from the LoBES survey \citep{Lynch_LoBES_2021}, centered around the EoR0 field (R.A. $0^h$, Dec $-27^\circ$), with the set of 128 deformed MWA beams described above. This simulation is performed at a 80kHz frequency resolution, over the 167-198 MHz band, and represents a fiducial ``measurement'' made by a realistically deformed and complex array. 

\subsection{Perfect \& Imperfect Calibration}
\label{ssec: per_imper_cal}
The fiducial simulation created in Section \ref{ssec: fid_sim} can be used to explore the effects of calibration errors introduced by the imperfect knowledge of beam models. We discriminate between two calibration scenarios below:

\begin{description}[style=unboxed,leftmargin=0cm]
    \item[\textbf{Perfect Calibration [$\mathbb{C}_{P}$]:}] In this case, a perfect understanding of our instrument is assumed, which is perfectly accounted for during calibration, along with a complete sky model. In particular, the set of deformed beam models used to generate the fiducial simulation in Section \ref{ssec: fid_sim} are used to generate the model visibilities for calibration ($V^{\mathrm{model}}$ from Eqn. \ref{eq: sky_cal}). This results in a perfect match between the fiducial simulation and the model visibilities used for calibration, leading to perfect calibration solutions.\\
    \item[\textbf{Imperfect Calibration [$\mathbb{C}_{I}$]:}] In this case, an incomplete understanding of our instrument is emulated by using a single, perfect (FEE) beam model to generate the the model visibilities for calibration. This scenario was chosen to mimic current interferometric calibration pipelines where varied or deformed beams are not considered. This case also uses a complete sky model, ensuring that any calibration errors arise purely from beam errors.
\end{description}

Following the application of these two calibration scenarios to our fiducial simulated data, a 2D (cylindrical) and 1D (spherical) power spectrum analysis is performed to quantify the effects of mismatches in instrumental and calibration beams on an EoR detection pipeline, described below.

\section{Power Spectrum}
\label{sec:ps}
The spatial power spectrum is designed to quantify spatial correlations in a cosmological field, and measures signal power as a function of spatial scale, \textit{k} ($h\mathrm{Mpc}^{-1}$). It can be defined as the Fourier transform of the two-point spatial correlation function: 

\begin{equation}
\begin{aligned}
P(|\vec{k}|)=\int_V\xi(\vec{r})e^{-2\pi i\vec{k}\cdot\vec{r}}d\vec{r},
\end{aligned}
\label{eq: ps_1}
\end{equation}

\noindent where $\xi(\vec{r})$ is the two-point spatial correlation function. The power spectrum can be estimated from the volume normalised Fourier transformed brightness temperature field, given an observing volume $\Omega$: 

\begin{equation}
\begin{aligned}
P(|\vec{k}|)\equiv\frac{1}{\Omega}\langle\Tilde{T}(k)^\dagger\Tilde{T}(k)\rangle.
\end{aligned}
\label{eq: ps_2}
\end{equation}

It's relevant to note that in an interferometer, the observing volume $\Omega$ is determined by the primary beam of each receiving element or tile. Given the nature of the satellite beam measurements made in \citealt{Chokshi_SAT_MAPS_2021}, we only consider changes to the shape of beam responses across the array in this work, and make no assertions regarding changing observing volumes. This is in contrast to the case of flagged or dead dipoles, which change both the beam shape as well as observed cosmological volumes \citep[see e.g.][]{joseph_bias_2018}.

Radio interferometers fundamentally sample Fourier modes across the spatial (angular) extent of the sky, captured by the measured interferometric visibilities (see Eq. \ref{eq: measurement_equation}): $\mathbf{u}\equiv(u, v) \mapsto k_\perp$. For a resonant line signal, such as the 21-cm line, line-of-sight Fourier modes can be mapped with the spectral channels: $\mathfrak{F}(f) = \eta \mapsto k_\parallel$. This mapping from measured interferometric visibility space $(u,v,f)$ to Fourier space $(u,v,\eta)$ leads to readily applicable expression for the power spectrum:

\begin{equation}
\begin{aligned}
P(|\vec{k}|)\equiv\frac{1}{\Omega}\langle\Tilde{V}(k)^\dagger\Tilde{V}(k)\rangle.
\end{aligned}
\label{eq: ps_3}
\end{equation}

In practice, multiple sets of measured visibilities are integrated coherently by gridding to a  discretised $uv$-plane, following a Fourier transform along the spectral axis which results in an $u,v,\eta$ data cube. This can now be squared to arrive at an unnormalised estimate of the cosmological power spectrum. Typically this orthogonal $k$-space is compressed to a 2D (cylindrically-averaged) and 1D (spherically-averaged) power spectra, where the former is used to isolate and diagnose foreground leakage and instrumental systematics, and the latter for cosmological measurements. The MWA EoR collaboration typically uses \texttt{CHIPS} - the Cosmological HI Power Spectrum estimator \citep{Trott_CHIPS_2016} and $\epsilon$ppsilon - Error Propagated Power Spectrum with Interleaved Observed Noise \citep{Barry_fhd_epp_2019} for power spectrum estimation. In this work we use \texttt{CHIPS} to perform our power spectrum analysis. 

\subsection{Foreground Contamination and Subtraction}

Galactic and extragalactic foregrounds dominate the faint cosmological EoR signal by up to five orders of magnitude (see Figure \ref{fig:slices}). To have any hope of detecting the EoR, extensive and accurate models of these foregrounds are necessary, including extended and bright sources such at Fornax A \citep{Line_Shapelets_2020}, diffuse emission \citep{byrne_diffuse_pol_2022}, the Galactic plane \citep{barry_gal_set_2024}, and the ubiquitous faint point-like extragalactic sources \citep{barry_calibration_2016}. A powerful discriminator between foreground flux and the background cosmological signal are their disparate spectral characteristics. The emission mechanisms of foreground sources are expected to be spectrally smooth \citep{Di_Matteo_Foregrounds_2002, oh_peng_foregrounds_2003}, while the 21-cm signal is anticipated to be uncorrelated on frequency scales larger than a MHz due to the topography of bubble formation and evolution as probed along the line-of-sight. 

The cylindrically-averaged 2D power spectrum is formed by collapsing the cartesian 3D $k$-space along the spatial extent $k_\perp=\sqrt{k_x^2+k_y^2}$, and the spectral or line-of-sight direction $k_\parallel$. In this space, spectrally smooth foregrounds components will dominate the low line-of-sight modes ($k_\parallel$) at all spatial modes perpendicular to the line-of-sight ($k_\perp$). We would thus expect a large region of this 2D $k$-space, above the low $k_\parallel$ modes, to be free of power from the intrinsic foreground components. Unfortunately, radio interferometers are chromatic - they exhibit frequency dependant responses in both their primary beams and their synthesized beam or Point Spread Function (PSF). This results in the well-documented ``foreground wedge'' caused by the mode-mixing of power from low $k_\parallel$ into larger $k_\parallel$ values \citep{datta_wedge_2010, morales_wedge_2012, vedantham_wedge_2012, parsons_wedge_2012, trott_wedge_2012, hazelton_wedge_2013, thyagarajan_wedge_2013, pober_wedge_2013, liu_wedge_2014a, liu_wedge_2014b, thyagarajan_wedge_2015}. This effect can also be considered to be the result of spectral structure being introduced to the otherwise spectrally smooth foregrounds by a chromatic instrumental response. The characteristic ``wedge'' shape of foreground mode-mixing arises from the fact that longer baselines (higher $k_\perp$) change more rapidly with frequency, resulting in faster spectral fluctuation which manifest as power at higher $k_\parallel$ modes.  

The area above the wedge is known as the ``EoR window'' and is expected to be contaminant free. The cosmological signal peaks at large scales, or small $k=\sqrt{k^2_\perp + k^2_\parallel}$, leading to an area of higher sensitivity in the lower left corner of the EoR window. This also means that $k$-modes within the wedge can have significantly more 21-cm power than those in the EoR window. The accurate subtraction of foreground flux from 21-cm data sets can enable the recovery of highly sensitive $k$-modes at the wedge-window boundary, boosting the significance of power spectrum measurements \citep{pober_next_gen_eor_2014, pober_foreground_removal_2016, beardsley_first_mwa_2016, cook_investigating_2022, barry_gal_set_2024}, and can theoretically put a statistical detection of the cosmological signal within reach of current generation experiments.

\section{Results}
\label{sec:results}

Following the two calibration scenarios described in Section \ref{ssec: per_imper_cal}, a systematic subtraction of foreground flux is performed to enable the recovery of $k$-modes around the edge of the EoR window. Of the 30,000 sources included in the fiducial simulation from Section \ref{ssec: fid_sim}, we generate model visibilities (using the relevant set of calibration beam models) with integrated apparent flux in 10$\%$ intervals (see Figure \ref{fig:srclist}), between the brightest 10\% to brightest 90\% to subtract from the two calibrated data sets. \texttt{CHIPS} is then used to calculate the 2D cylindrical-averaged power spectrum, and a 1D spherically-averaged power spectrum within the EoR window.

\subsection{2D Power Spectrum}

\begin{figure}
    \includegraphics[width=\columnwidth]{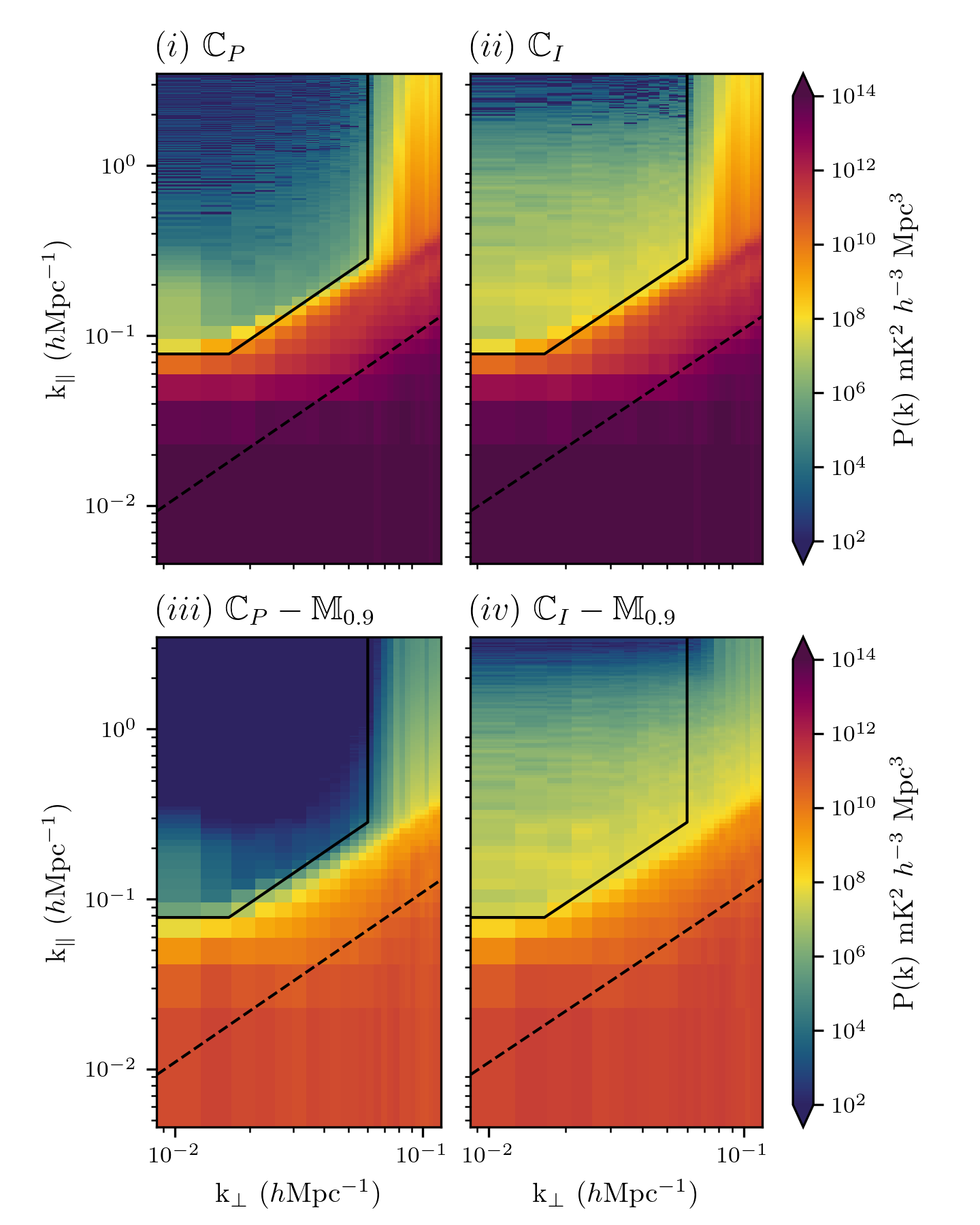}
    \caption{Cylindrical-averaged power spectra of the two calibration scenarios described in Section \ref{ssec: per_imper_cal}. The left column (panels \textit{i, iii}) represents perfect calibration where the varied beam models are accounted for during calibration. The right column (panels \textit{ii, iv}) represent imperfect calibration when a simple and incomplete instrumental model is used for calibration. The bottom row (panels \textit{iii, iv}) are identical to the top row (panels \textit{i, ii}) except that a 90$\%$ of the brightest foreground flux have been subtracted. The top left region of each panel is the EoR window where a search for the cosmological signal can be performed.}
    \label{fig:2dps}
\end{figure}

The 2D or cylindrically-averaged power spectrum is the compressed parameter space where line-of-sight modes ($k_\parallel$) and those in the orthogonal plane of the sky ($k_\perp$) are separated, making it an ideal space to observe and understand the complex effects of foreground-instrumental coupling \citep{pober_foreground_removal_2016}. Figure \ref{fig:2dps} displays the 2D power spectra of the two calibration scenarios described in Section \ref{ssec: per_imper_cal} before and after the majority of foreground flux has been subtracted. The dashed lines in Figure \ref{fig:2dps} represent the full-width half-max of the MWA primary beam response, while the black contours in the upper left corner of each panel represent the EoR window above the horizon.

The first two panels ($i,\;ii$) of Figure \ref{fig:2dps} contain flux from all 30,000 sources included in the simulation, with key differences occurring in the EoR window in the top left. The EoR window of the perfect calibration case ($\mathbb{C}_P$: panel ($i$)) has much less foreground power than the imperfect calibration case ($\mathbb{C}_I$: panel ($ii$)), by a factor of approximately 100. This excess foreground power present in the EoR window can be completely attributed to the mismatch between the set of measurement beam models (used to create the fiducial simulations in Section \ref{ssec: fid_sim}) and the single perfect beam model used during calibration. 

We subsequently subtract a sky-model, generated with the relevant set of beams, containing 90$\%$ of the brightest apparent flux (see Figure \ref{fig:srclist}) from each calibrated data set results. This results in an anticipated reduction of power within the foreground-wedge (lower sections of panels $iii,\;iv$), but unexpected behaviour within the EoR window. In the perfect calibration case ($\mathbb{C}_P - \mathbb{M}_{0.9}$: panel ($iii$)), the EoR window has significantly reduced power, while in the imperfect calibration case ($\mathbb{C}_I - \mathbb{M}_{0.9}$: panel ($iv$)), the EoR window power has remained essentially the same. The difference in EoR window power after foreground subtraction has now widened to be greater than a factor of 10,000, reaching levels significantly below the expected EoR in the perfect calibration case (panel ($iv$)). 

This implies that spectral structure introduced into the calibration solutions by the mismatch between the set of instrumental and single calibration beam leads to mode mixing from low $k_\parallel$ modes to high $k_\parallel$ well beyond the expected foreground wedge. It also demonstrates that this excess beam-based chromaticity introduces power to the EoR window which cannot be mitigated by simply subtracting partial models of the foregrounds.

 \subsection{1D Power Spectrum}

Spherically averaging the $k$-modes within the EoR window leads to a 1D power spectrum typically assumed to be free of foreground power which can then be used to make cosmological measurements. In this work, we use the 1D power spectrum to quantify the extent of foreground spectral leakage into the EoR window caused by the differing instrumental and calibration beams. 

\begin{figure}
    \includegraphics[width=\columnwidth]{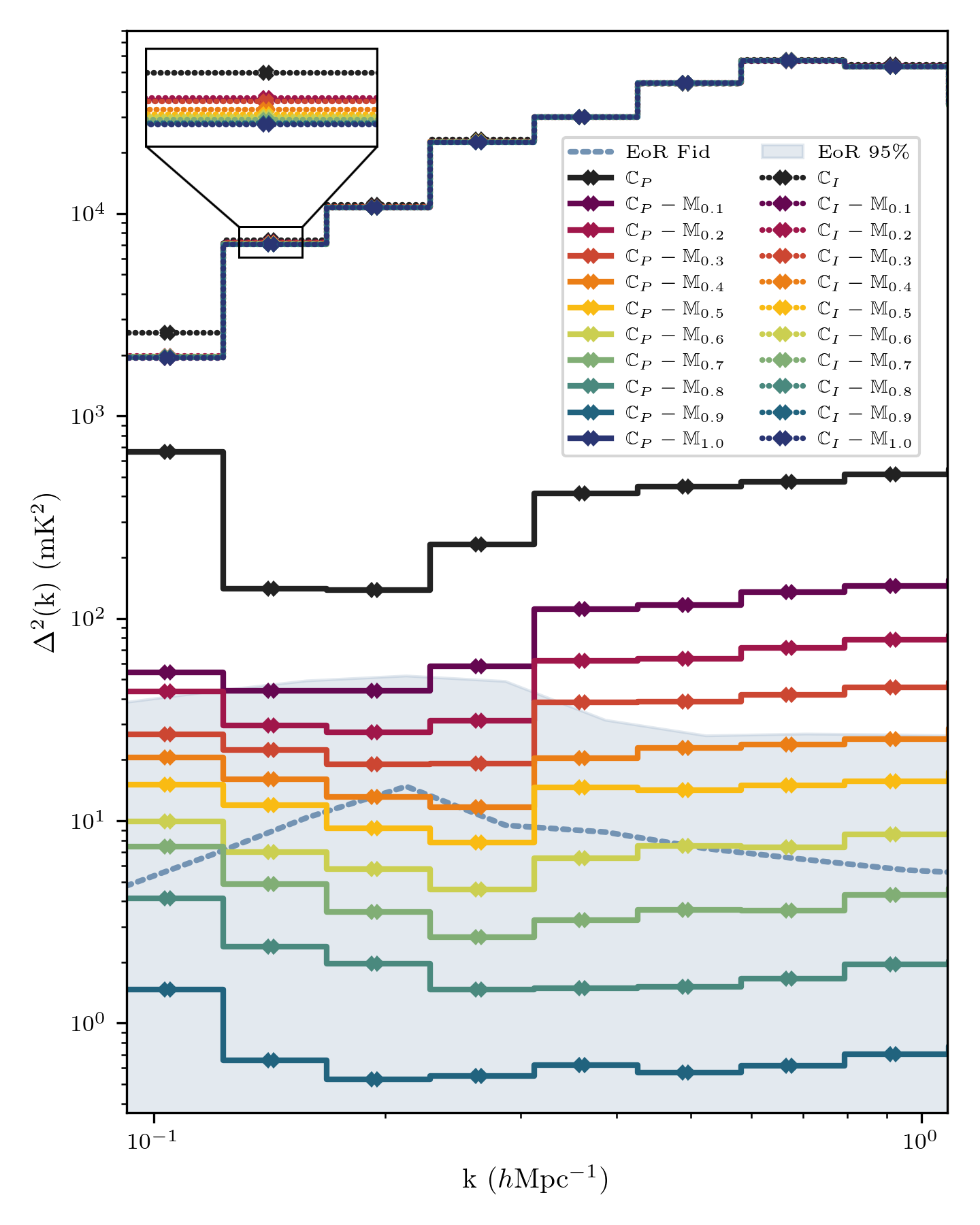}
    \caption{Spherically-averaged power spectra within the EoR window. The dashed lines represent the imperfect calibration ($\mathbb{C}_I$) scenario, while the solid lines represent the perfect calibration ($\mathbb{C}_P$) case. The coloured lines represent a systematic subtraction of apparent foreground flux, in intervals between 10\% to 100\%  The grey dotted line is the fiducial EoR level while the shaded region represents the 95\% confidence limits. Note that all the dashed lines from the imperfect calibration scenario lie practically on top of one another, and do not change significantly after subtracting foreground flux models.}
    \label{fig:1dps}
\end{figure}

In Figure \ref{fig:1dps} the grey dotted line and shaded regions denote the power level of an EoR model and its 95\% confidence limits \citep{barry_eor_lim_2019b, Greig_large_cosomo_2022} - which are used as a reference to compare levels of beam-based spectral leakage. Only when foreground leakage into the EoR window is below the EoR level and thus a sub-dominant systematic, is there any hope of a direct measurement of the cosmological signal. The different colours in Figure~\ref{fig:1dps} denote varied levels of foreground subtraction from the brightest 10\% in apparent flux to a complete 100\% of all sources in 10\% integrated flux bins (see Figure \ref{fig:srclist}). The solid and dashed lines represent the perfect and imperfect calibration cases respectively.  

In the imperfect calibration case (dashed lines in Figure \ref{fig:1dps}) when an incomplete model of the telescope (i.e. a single beam model) is assumed during calibration, the resultant spectral structure introduced into the calibration solutions leads to foreground spectral leakage over a 1000 times our fiducial EoR model between $k$=0.1 and $k$=1 $h\mathrm{Mpc}^{-1}$. This foreground leakage into the EoR window is not appreciably reduced by subtracting models of foreground sources (dashed lines in Figure \ref{fig:1dps} lie practically on top of one another), demonstrating that the excess chromaticity introduced by beam-based calibration errors results in mode mixing beyond the well characterised foreground-wedge effect caused by instrumental chromaticity.

If an accurate instrumental model is used during calibration, as demonstrated by the perfect calibration scenario (solid lines in Figure \ref{fig:1dps}), systematically subtracting models of the brightest apparent flux reduces spectral leakage into the EoR window till it is a sub-dominant effect. In fact, the solid navy blue line which represents a complete subtraction of foreground flux ($\mathbf{C}_P-\mathbb{M}_{1.0}$) during perfect calibration lies at the $\sim10^{-20}$ $\mathrm{mK}^2$ level far below the bottom of the y-axis in Figure. \ref{fig:1dps}, and is numerically insignificant. This demonstrates that in the perfect calibration scenario, all foreground flux which is known can be subtracted from the EoR window, in contrast to the imperfect calibration scenario where a fundamental spectral leakage imprint remains in the EoR window despite the subtraction of sky-model flux. 

A pertinent question to consider is why there is any power in the perfect calibration case prior to any flux is subtraction ($\mathbb{C}_P$: black solid line in Figure \ref{fig:1dps}), since all the foreground flux is expected to be contained in the foreground-wedge. We primarily attribute this to excess chromaticity from the implementation of the FEE beam model, but can also arise from the bandpass, decoherence due to frequency smearing, and other unidentified analysis or instrumental systematics. Any excess chromaticity leads to mode mixing of power from the foreground-wedge into the EoR window, which is then measured in the spherically-averaged power spectrum.

\subsection{Spectral Structure in Calibration Solutions}

\begin{figure*}
    \includegraphics[width=\textwidth]{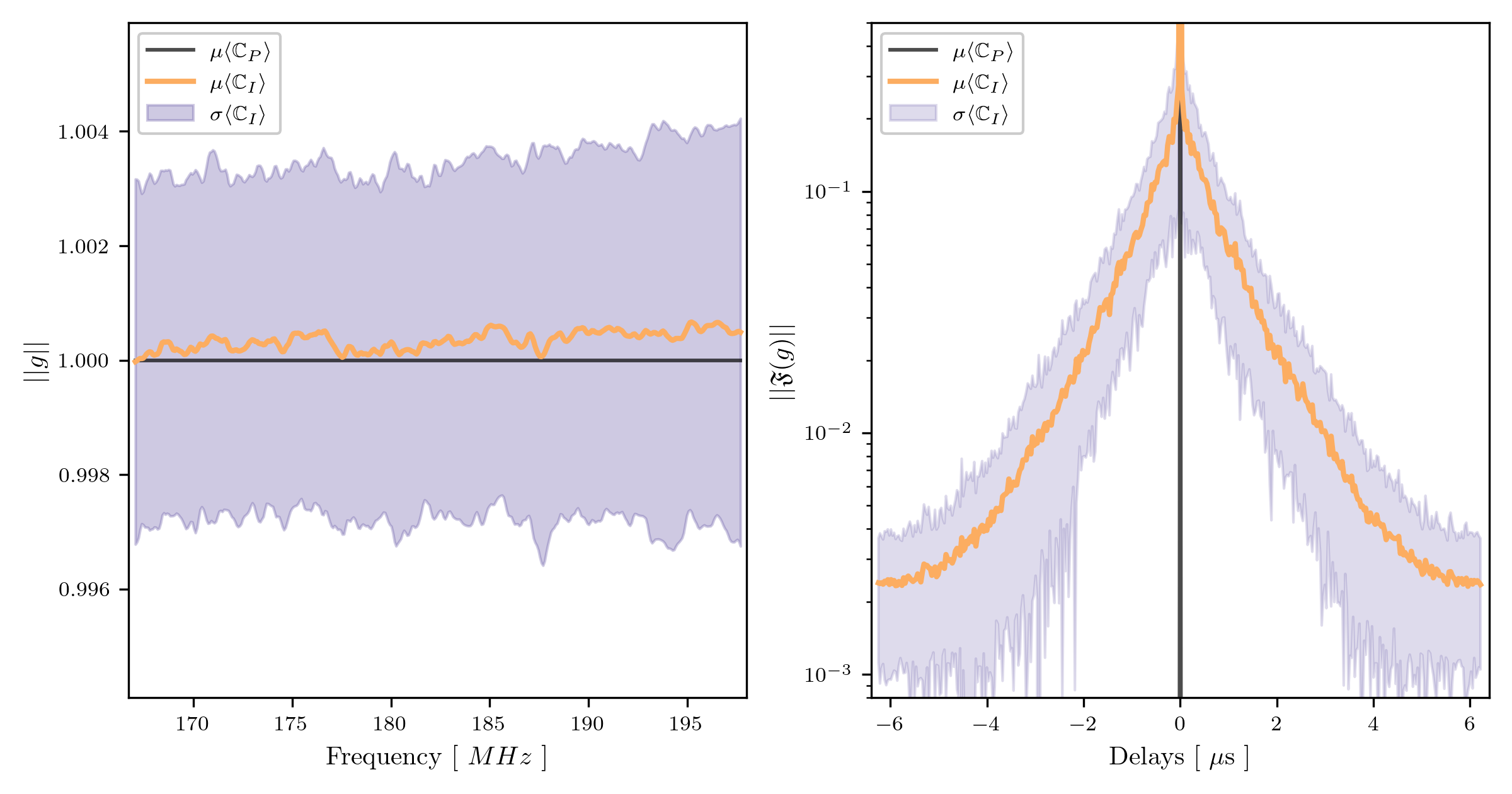}
    \caption{Gain amplitudes of calibration solutions are shown in the left panel, with the black line representing perfect calibration averaged over tiles ($\mu\langle\mathbb{C}_P\rangle$), while the yellow line is the antenna averaged gain amplitudes in the imperfect calibration scenario ($\mu\langle\mathbb{C}_I\rangle$) with the purple region enclosing 68\% of values ($\sigma\langle\mathbb{C}_I\rangle$). The right panel is the Fourier transform along frequency of gain amplitudes, and plot the results as a function of delay. The black line is perfect calibration, while the yellow line represent the antenna averaged quantity and purple region encloses 68\% of values from the imperfect calibration scenario.}
    \label{fig:cal_del}
\end{figure*}

The 2D and 1D power spectra clearly demonstrate the effects of beam-based calibration errors, yet it is instructive to explore the raw calibration solutions obtained in Section \ref{ssec: per_imper_cal}, where beam-based spectral leakage initially originates.  

The fiducial simulation from Section \ref{ssec: fid_sim} was noiseless, and used a sky catalogue of 30,000 sources to generate visibilities which emulated a measurement by the MWA with a set of deformed beam models ($V^{\mathrm{data}}$). While thermal noise can be a significant systematic in single observations, for a temporally stable instrument such as the MWA, noise in calibration solutions has been shown to incoherently average \citep{Barry_fhd_epp_2019}. In the perfect calibration scenario, an identical set of deformed beams and sky catalogue are used to generate model visibilities ($V^{\mathrm{model}}$) for the calibration minimisation process (see Equation \ref{eq: sky_cal}). In the absence of noise, the fact that the data and model visibilities are identical leads to gain solutions which are unity within double precision across the frequency band (black line in the left panel of Figure \ref{fig:cal_del}). Adding thermal noise to the simulations would introduce uncertainty in the calibration solutions leading to a deviation for unity described above. This has the potential to introduce spectral leakage into the EoR window, even in simulations using a single perfect beam model. Further investigations along this line are left for future works.

In the imperfect calibration scenario, a single perfect beam model is used to generate the model visibilities ($V^{\mathrm{model}}$) with the original sky catalogue. During the calibration process, the mismatch between the data and model visibilities lead to non-unity gain solutions as a function of frequency. This frequency structure is the root cause of foreground leakage from the EoR wedge into the EoR window observed in the power spectra, and can be solely attributed to an incomplete representation of the instrument (single beam model instead of set of deformed beam models) during the calibration process. The yellow solid line in the left panel of Figure \ref{fig:cal_del} represent the antenna-averaged calibration gain amplitudes to visualise any common spectral structure, while the purple region encloses 68\% of values. 

To gauge the spectral structure within calibration solutions, we perform a Fourier transform across frequency, which decomposes calibration error amplitudes as a function of delay modes. In the perfect calibration scenario, this results in a delta function at a delay of zero, and any deviation from this would lead to excess calibration chromaticity, resulting in mode mixing from the foreground-wedge into the EoR window. The right panel of the Figure displays the Fourier transform of the calibration gain amplitudes, with the green line being the antenna-averaged quantity, while the blue region again encloses 68\% of values. 

\section{Conclusions \& Next Steps}
\label{sec:conclusions}


This work explores the impact that imperfect and varied beams across a radio interferometer could have on EoR power spectrum measurements. We demonstrate how incomplete representations of varied and complex beams during calibration can lead to the leakage of foreground power into modes sensitive to the cosmological signal. This leads to contamination beyond the well-known foreground wedge into the EoR window, which is typically assumed to be free of foreground contaminants, at levels which exceed the expected EoR level by factors greater than $\sim$1000 between $k$=0.1-1 $h\mathrm{Mpc}^{-1}$. We also demonstrate how this effect is not improved by subtracting models and foreground sources, and necessitates the inclusion of validated and measured beam models in calibration frameworks.

Appreciable differences have been measured between cutting-edge electromagnetic Fully Embedded Element (FEE) MWA beam model \citep{Sutinjo_FEE_2015} and in-situ measurements using satellites \citep{Line_ORBCOMM_2018, Chokshi_SAT_MAPS_2021}. These effects are predominantly measured as deformations in beam sidelobes, and are attributed to a variety of environmental factors. In Section \ref{sec:sat_maps}, we develop a method of leveraging the 16 dipole gain parameters, natively used to weight the contribution of each dipole to the summed MWA tile response, to deform the FEE beam model to best match satellite beam maps from \citet{Chokshi_SAT_MAPS_2021}. 

In Section \ref{sec:cali}, we develop a physically motivated model to simulate a full 128-tile MWA array composed of realistically deformed beams based on the 14 dual polarization maps available from \citet{Chokshi_SAT_MAPS_2021}. Using 30,000 complex sources from the LoBES catalog \citep{Lynch_LoBES_2021}, we create a noiseless fiducial simulation using the set of deformed beams, which emulates a real measurement with the MWA. We now calibrate our fiducial simulation using two strategies; perfect calibration where a complete understanding of the instrument is assumed by using the set of deformed beams during calibration, or imperfect calibration where a single FEE beam model is used to emulate currently accepted calibration strategies which do not account for beam variations across the radio interferometer. Using the spatial power spectrum described in Section \ref{sec:ps}, we investigate the effects of beam-induced calibration errors on the prospects of recovering an EoR signal in Section \ref{sec:results} (see Figures \ref{fig:2dps} \& \ref{fig:1dps}).

Our work demonstrates that \textbf{including physically motivated beam models during calibration has the potential to reduce foreground spectral leakage into the EoR window by factors greater that 1000, which could potentially put a statistical detection of the cosmological signal within grasp}. We outline the necessary step required to make this technique applicable to real data below:

\begin{itemize}[wide, labelwidth=!,itemindent=!,labelindent=0pt, leftmargin=0em, itemsep=1.4mm]
    \item Satellite beam maps of each station in the radio interferometer will be a crucial first step. The satellite backend developed in \citet{Line_ORBCOMM_2018, Chokshi_SAT_MAPS_2021} and the \texttt{EMBERS} analysis pipeline \citep{Chokshi_EMBERS_2020} have provided excellent all-sky maps at 137MHz at a very reasonable expense. 
    \item A key question which has not been addressed by this work is the fact that the satellite maps by \citet{Chokshi_SAT_MAPS_2021} span a 1.4~MHz bandwidth, and are relatively narrowband in comparison to the observing bandwidth of $\sim30$~MHz. Thus the beam deformation model described in Section \ref{sec:sat_maps} is generated from narrowband data and applied across the broader observing band. The validity of this approach must be validated and augmented using a drone-based beam measurement system \citep{Chang_drone_2015, Jacobs_drones_2017, Bolli_drones_lofar_2018, Ninni_drones_lofar_2020, Paonessa_drones_ska_2020, Herman_Drone_2024}.
    \item The 16 parameter beam deformation model developed in this work was physically motivated by the aperture array design of MWA tiles, and could be modified to be applicable to telescopes such as LOFAR, NenuFAR, or the future SKA-Low. Unfortunately, telescopes such as HERA which employ parabolic dishes as their intereferometric elements will require new innovative models such as those developed by \citet{wilensky_beam_2024}. 
\end{itemize}

We have demonstrated how a mismatch between the complex set of instrumental beams and the beam assumed during calibration can lead to the introduction of artificial spectral structure into calibration solutions which results in foreground leakages beyond the foreground wedge and into the EoR window. While we have shown that this beam-based calibration leakage can be mitigated by the inclusion of more accurate representations of instrumental beam models into calibration frameworks, it it not necessarily the only solution. While beyond the scope of this work, we leave the investigation of direction-dependant calibration, delay filtering, and regularised calibration solutions for future works.

\section*{Acknowledgements}

We are grateful to Miguel F. Morales for insightful discussions.

This research was supported by the Australian Research Council Centre of Excellence for All Sky Astrophysics in 3 Dimensions (ASTRO 3D), through project number CE170100013. Part of this work was supported by the Melbourne Research Scholarship from the University of Melbourne. Part of this work was supported by the Australian Research Council Discovery Early Career Researcher Award (DECRA) through project number DE240101377. This work was supported by resources awarded under Astronomy Australia Ltd’s merit allocation scheme on the OzSTAR national facility at Swinburne University of Technology. OzSTAR is funded by Swinburne University of Technology and the National Collaborative Research Infrastructure Strategy (NCRIS). This work was supported by resources provided by the Pawsey Supercomputing Centre with funding from the Australian Government and the Government of Western Australia. This scientific work makes use of the Inyarrimanha Ilgari Bundara Murchison Radio-Astronomy Observatory, operated by CSIRO. We acknowledge the Wajarri Yamatji people as the traditional owners of the Observatory site. The International Centre for Radio Astronomy Research (ICRAR) is a Joint Venture of Curtin University and The University of Western Australia, funded by the Western Australian State government.

\section*{Data Availability}

The data underlying this article will be shared on reasonable
request to the corresponding author.



\bibliographystyle{mnras}
\bibliography{paper} 

\newcommand{\SortNoop}[1]{}
\begin{thebibliography}{}
\makeatletter
\relax
\def\mn@urlcharsother{\let\do\@makeother \do\$\do\&\do\#\do\^\do\_\do\%\do\~}
\def\mn@doi{\begingroup\mn@urlcharsother \@ifnextchar [ {\mn@doi@} {\mn@doi@[]}}
\def\mn@doi@[#1]#2{\def\@tempa{#1}\ifx\@tempa\@empty \href {http://dx.doi.org/#2} {doi:#2}\else \href {http://dx.doi.org/#2} {#1}\fi \endgroup}
\def\mn@eprint#1#2{\mn@eprint@#1:#2::\@nil}
\def\mn@eprint@arXiv#1{\href {http://arxiv.org/abs/#1} {{\tt arXiv:#1}}}
\def\mn@eprint@dblp#1{\href {http://dblp.uni-trier.de/rec/bibtex/#1.xml} {dblp:#1}}
\def\mn@eprint@#1:#2:#3:#4\@nil{\def\@tempa {#1}\def\@tempb {#2}\def\@tempc {#3}\ifx \@tempc \@empty \let \@tempc \@tempb \let \@tempb \@tempa \fi \ifx \@tempb \@empty \def\@tempb {arXiv}\fi \@ifundefined {mn@eprint@\@tempb}{\@tempb:\@tempc}{\expandafter \expandafter \csname mn@eprint@\@tempb\endcsname \expandafter{\@tempc}}}

\bibitem[\protect\citeauthoryear{Barry, Hazelton, Sullivan, Morales  \& Pober}{Barry et~al.}{2016}]{barry_calibration_2016}
Barry N.,  Hazelton B.,  Sullivan I.,  Morales M.~F.,   Pober J.~C.,  2016, \mn@doi [Monthly Notices of the Royal Astronomical Society] {10.1093/mnras/stw1380}, 461, 3135

\bibitem[\protect\citeauthoryear{{Barry}, {Beardsley}, {Byrne}, {Hazelton}, {Morales}, {Pober}  \& {Sullivan}}{{Barry} et~al.}{2019a}]{Barry_fhd_epp_2019}
{Barry} N.,  {Beardsley} A.~P.,  {Byrne} R.,  {Hazelton} B.,  {Morales} M.~F.,  {Pober} J.~C.,   {Sullivan} I.,  2019a, \mn@doi [\pasa] {10.1017/pasa.2019.21}, \href {https://ui.adsabs.harvard.edu/abs/2019PASA...36...26B} {36, e026}

\bibitem[\protect\citeauthoryear{{Barry} et~al.,}{{Barry} et~al.}{2019b}]{barry_eor_lim_2019b}
{Barry} N.,  et~al., 2019b, \mn@doi [\apj] {10.3847/1538-4357/ab40a8}, \href {https://ui.adsabs.harvard.edu/abs/2019ApJ...884....1B} {884, 1}

\bibitem[\protect\citeauthoryear{{Barry}, {Line}, {Lynch}, {Kriele}  \& {Cook}}{{Barry} et~al.}{2024}]{barry_gal_set_2024}
{Barry} N.,  {Line} J.~L.~B.,  {Lynch} C.~R.,  {Kriele} M.,   {Cook} J.,  2024, \mn@doi [\apj] {10.3847/1538-4357/ad2e9b}, \href {https://ui.adsabs.harvard.edu/abs/2024ApJ...964..158B} {964, 158}

\bibitem[\protect\citeauthoryear{{Beardsley} et~al.,}{{Beardsley} et~al.}{2016}]{beardsley_first_mwa_2016}
{Beardsley} A.~P.,  et~al., 2016, \mn@doi [\apj] {10.3847/1538-4357/833/1/102}, \href {https://ui.adsabs.harvard.edu/abs/2016ApJ...833..102B} {833, 102}

\bibitem[\protect\citeauthoryear{{Bolli}, {Pupillo}, {Paonessa}, {Virone}, {Wijnholds}  \& {Lingua}}{{Bolli} et~al.}{2018}]{Bolli_drones_lofar_2018}
{Bolli} P.,  {Pupillo} G.,  {Paonessa} F.,  {Virone} G.,  {Wijnholds} S.~J.,   {Lingua} A.~M.,  2018, \mn@doi [IEEE Antennas and Wireless Propagation Letters] {10.1109/LAWP.2018.2805999}, 17, 613

\bibitem[\protect\citeauthoryear{Bowman et~al.,}{Bowman et~al.}{2007}]{bowman_field_2007}
Bowman J.~D.,  et~al., 2007, \mn@doi [The Astronomical Journal] {10.1086/511068}, 133, 1505

\bibitem[\protect\citeauthoryear{Byrne et~al.,}{Byrne et~al.}{2019}]{byrne_fundamental_2019}
Byrne R.,  et~al., 2019, \mn@doi [The Astrophysical Journal] {10.3847/1538-4357/ab107d}, 875, 70

\bibitem[\protect\citeauthoryear{{Byrne}, {Morales}, {Hazelton}, {Sullivan}, {Barry}, {Lynch}, {Line}  \& {Jacobs}}{{Byrne} et~al.}{2022}]{byrne_diffuse_pol_2022}
{Byrne} R.,  {Morales} M.~F.,  {Hazelton} B.,  {Sullivan} I.,  {Barry} N.,  {Lynch} C.,  {Line} J. L.~B.,   {Jacobs} D.~C.,  2022, \mn@doi [\mnras] {10.1093/mnras/stab3276}, \href {https://ui.adsabs.harvard.edu/abs/2022MNRAS.510.2011B} {510, 2011}

\bibitem[\protect\citeauthoryear{{Chang}, {Monstein}, {Refregier}, {Amara}, {Glauser}  \& {Casura}}{{Chang} et~al.}{2015}]{Chang_drone_2015}
{Chang} C.,  {Monstein} C.,  {Refregier} A.,  {Amara} A.,  {Glauser} A.,   {Casura} S.,  2015, \mn@doi [\pasp] {10.1086/683467}, \href {https://ui.adsabs.harvard.edu/abs/2015PASP..127.1131C} {127, 1131}

\bibitem[\protect\citeauthoryear{Chokshi, b. Line  \& McKinley}{Chokshi et~al.}{2020}]{Chokshi_EMBERS_2020}
Chokshi A.,  b. Line J.~L.,   McKinley B.,  2020, \mn@doi [Journal of Open Source Software] {10.21105/joss.02629}, 5, 2629

\bibitem[\protect\citeauthoryear{Chokshi, Line, Barry, Ung, Kenney, McPhail, Williams  \& Webster}{Chokshi et~al.}{2021}]{Chokshi_SAT_MAPS_2021}
Chokshi A.,  Line J. L.~B.,  Barry N.,  Ung D.,  Kenney D.,  McPhail A.,  Williams A.,   Webster R.~L.,  2021, \mn@doi [Monthly Notices of the Royal Astronomical Society] {10.1093/mnras/stab156}, 502, 1990–2004

\bibitem[\protect\citeauthoryear{Choudhuri, Bull  \& Garsden}{Choudhuri et~al.}{2021}]{choudhuri_patterns_2021}
Choudhuri S.,  Bull P.,   Garsden H.,  2021, \mn@doi [Monthly Notices of the Royal Astronomical Society] {10.1093/mnras/stab1795}, 506, 2066

\bibitem[\protect\citeauthoryear{Cook, Trott  \& Line}{Cook et~al.}{2022}]{cook_investigating_2022}
Cook J.~H.,  Trott C.~M.,   Line J. L.~B.,  2022, \mn@doi [MNRAS] {10.1093/mnras/stac1330}, 514, 790

\bibitem[\protect\citeauthoryear{{Datta}, {Bowman}  \& {Carilli}}{{Datta} et~al.}{2010}]{datta_wedge_2010}
{Datta} A.,  {Bowman} J.~D.,   {Carilli} C.~L.,  2010, \mn@doi [\apj] {10.1088/0004-637X/724/1/526}, \href {https://ui.adsabs.harvard.edu/abs/2010ApJ...724..526D} {724, 526}

\bibitem[\protect\citeauthoryear{DeBoer et~al.,}{DeBoer et~al.}{2017}]{deboer_hydrogen_2017}
DeBoer D.~R.,  et~al., 2017, \mn@doi [Publications of the Astronomical Society of the Pacific] {10.1088/1538-3873/129/974/045001}, 129, 045001

\bibitem[\protect\citeauthoryear{{Di Matteo}, {Perna}, {Abel}  \& {Rees}}{{Di Matteo} et~al.}{2002}]{Di_Matteo_Foregrounds_2002}
{Di Matteo} T.,  {Perna} R.,  {Abel} T.,   {Rees} M.~J.,  2002, \mn@doi [\apj] {10.1086/324293}, \href {https://ui.adsabs.harvard.edu/abs/2002ApJ...564..576D} {564, 576}

\bibitem[\protect\citeauthoryear{Eastwood et~al.,}{Eastwood et~al.}{2019}]{eastwood_21_2019}
Eastwood M.~W.,  et~al., 2019, \mn@doi [The Astronomical Journal] {10.3847/1538-3881/ab2629}, 158, 84

\bibitem[\protect\citeauthoryear{Foreman-Mackey, Hogg, Lang  \& Goodman}{Foreman-Mackey et~al.}{2013}]{Foreman_Mackey_EMCEE_2013}
Foreman-Mackey D.,  Hogg D.~W.,  Lang D.,   Goodman J.,  2013, \mn@doi [Publications of the Astronomical Society of the Pacific] {10.1086/670067}, 125, 306–312

\bibitem[\protect\citeauthoryear{{G{\'o}rski}, {Hivon}, {Banday}, {Wandelt}, {Hansen}, {Reinecke}  \& {Bartelmann}}{{G{\'o}rski} et~al.}{2005}]{Gorski_healpix_2005}
{G{\'o}rski} K.~M.,  {Hivon} E.,  {Banday} A.~J.,  {Wandelt} B.~D.,  {Hansen} F.~K.,  {Reinecke} M.,   {Bartelmann} M.,  2005, \mn@doi [\apj] {10.1086/427976}, \href {https://ui.adsabs.harvard.edu/abs/2005ApJ...622..759G} {622, 759}

\bibitem[\protect\citeauthoryear{{Greig}, {Wyithe}, {Murray}, {Mutch}  \& {Trott}}{{Greig} et~al.}{2022}]{Greig_large_cosomo_2022}
{Greig} B.,  {Wyithe} J. S.~B.,  {Murray} S.~G.,  {Mutch} S.~J.,   {Trott} C.~M.,  2022, \mn@doi [\mnras] {10.1093/mnras/stac2506}, \href {https://ui.adsabs.harvard.edu/abs/2022MNRAS.516.5588G} {516, 5588}

\bibitem[\protect\citeauthoryear{{\SortNoop{Haarlem}}van~Haarlem et~al.,}{{\SortNoop{Haarlem}}van~Haarlem et~al.}{2013}]{van_haarlem_lofar:_2013}
{\SortNoop{Haarlem}}van~Haarlem M.~P.,  et~al., 2013, \mn@doi [Astronomy \& Astrophysics] {10.1051/0004-6361/201220873}, 556, A2

\bibitem[\protect\citeauthoryear{{Hazelton}, {Morales}  \& {Sullivan}}{{Hazelton} et~al.}{2013}]{hazelton_wedge_2013}
{Hazelton} B.~J.,  {Morales} M.~F.,   {Sullivan} I.~S.,  2013, \mn@doi [\apj] {10.1088/0004-637X/770/2/156}, \href {https://ui.adsabs.harvard.edu/abs/2013ApJ...770..156H} {770, 156}

\bibitem[\protect\citeauthoryear{{Herman} et~al.,}{{Herman} et~al.}{2024}]{Herman_Drone_2024}
{Herman} L.,  et~al., 2024, \mn@doi [arXiv e-prints] {10.48550/arXiv.2407.00856}, \href {https://ui.adsabs.harvard.edu/abs/2024arXiv240700856H} {p. arXiv:2407.00856}

\bibitem[\protect\citeauthoryear{{Hinton}}{{Hinton}}{2016}]{Hinton_chain_consumer_2016}
{Hinton} S.~R.,  2016, \mn@doi [The Journal of Open Source Software] {10.21105/joss.00045}, \href {http://adsabs.harvard.edu/abs/2016JOSS....1...45H} {1, 00045}

\bibitem[\protect\citeauthoryear{{Jacobs} et~al.,}{{Jacobs} et~al.}{2017}]{Jacobs_drones_2017}
{Jacobs} D.~C.,  et~al., 2017, \mn@doi [\pasp] {10.1088/1538-3873/aa56b9}, \href {https://ui.adsabs.harvard.edu/abs/2017PASP..129c5002J} {129, 035002}

\bibitem[\protect\citeauthoryear{{Jeli{\'c}} et~al.,}{{Jeli{\'c}} et~al.}{2008}]{jelic_firegrounds_2008}
{Jeli{\'c}} V.,  et~al., 2008, \mn@doi [\mnras] {10.1111/j.1365-2966.2008.13634.x}, \href {https://ui.adsabs.harvard.edu/abs/2008MNRAS.389.1319J} {389, 1319}

\bibitem[\protect\citeauthoryear{Joseph, Trott  \& Wayth}{Joseph et~al.}{2018}]{joseph_bias_2018}
Joseph R.~C.,  Trott C.~M.,   Wayth R.~B.,  2018, \mn@doi [The Astronomical Journal] {10.3847/1538-3881/aaec0b}, 156, 285

\bibitem[\protect\citeauthoryear{Joseph, Trott, Wayth  \& Nasirudin}{Joseph et~al.}{2020}]{joseph_calibration_2020}
Joseph R.~C.,  Trott C.~M.,  Wayth R.~B.,   Nasirudin A.,  2020, \mn@doi [Monthly Notices of the Royal Astronomical Society] {10.1093/mnras/stz3375}, 492, 2017

\bibitem[\protect\citeauthoryear{Kim et~al.,}{Kim et~al.}{2022}]{kim_impact_2022}
Kim H.,  et~al., 2022, \mn@doi [The Astrophysical Journal] {10.3847/1538-4357/ac9eaf}, 941, 207

\bibitem[\protect\citeauthoryear{Koopmans et~al.,}{Koopmans et~al.}{2015}]{koopmans_cosmic_2015}
Koopmans L.,  et~al., 2015, Advancing Astrophysics with the Square Kilometre Array (AASKA14), p.~1

\bibitem[\protect\citeauthoryear{Line et~al.,}{Line et~al.}{2018}]{Line_ORBCOMM_2018}
Line J. L.~B.,  et~al., 2018, \mn@doi [Publications of the Astronomical Society of Australia] {10.1017/pasa.2018.30}, 35, e045

\bibitem[\protect\citeauthoryear{Line et~al.,}{Line et~al.}{2020}]{Line_Shapelets_2020}
Line J. L.~B.,  et~al., 2020, \mn@doi [Publications of the Astronomical Society of Australia] {10.1017/pasa.2020.18}, 37

\bibitem[\protect\citeauthoryear{{Liu}, {Parsons}  \& {Trott}}{{Liu} et~al.}{2014a}]{liu_wedge_2014a}
{Liu} A.,  {Parsons} A.~R.,   {Trott} C.~M.,  2014a, \mn@doi [\prd] {10.1103/PhysRevD.90.023018}, \href {https://ui.adsabs.harvard.edu/abs/2014PhRvD..90b3018L} {90, 023018}

\bibitem[\protect\citeauthoryear{{Liu}, {Parsons}  \& {Trott}}{{Liu} et~al.}{2014b}]{liu_wedge_2014b}
{Liu} A.,  {Parsons} A.~R.,   {Trott} C.~M.,  2014b, \mn@doi [\prd] {10.1103/PhysRevD.90.023019}, \href {https://ui.adsabs.harvard.edu/abs/2014PhRvD..90b3019L} {90, 023019}

\bibitem[\protect\citeauthoryear{{Lynch} et~al.,}{{Lynch} et~al.}{2021}]{Lynch_LoBES_2021}
{Lynch} C.~R.,  et~al., 2021, \mn@doi [\pasa] {10.1017/pasa.2021.50}, \href {https://ui.adsabs.harvard.edu/abs/2021PASA...38...57L} {38, e057}

\bibitem[\protect\citeauthoryear{Mellema et~al.,}{Mellema et~al.}{2013}]{mellema_reionization_2013}
Mellema G.,  et~al., 2013, \mn@doi [Experimental Astronomy] {10.1007/s10686-013-9334-5}, 36, 235

\bibitem[\protect\citeauthoryear{Mitchell, Greenhill, Wayth, Sault, Lonsdale, Cappallo, Morales  \& Ord}{Mitchell et~al.}{2008}]{Mitchell_RTS_2008}
Mitchell D.~A.,  Greenhill L.~J.,  Wayth R.~B.,  Sault R.~J.,  Lonsdale C.~J.,  Cappallo R.~J.,  Morales M.~F.,   Ord S.~M.,  2008, IEEE Journal of Selected Topics in Signal Processing, 2, 707

\bibitem[\protect\citeauthoryear{{Morales}, {Hazelton}, {Sullivan}  \& {Beardsley}}{{Morales} et~al.}{2012}]{morales_wedge_2012}
{Morales} M.~F.,  {Hazelton} B.,  {Sullivan} I.,   {Beardsley} A.,  2012, \mn@doi [\apj] {10.1088/0004-637X/752/2/137}, \href {https://ui.adsabs.harvard.edu/abs/2012ApJ...752..137M} {752, 137}

\bibitem[\protect\citeauthoryear{{Munshi} et~al.,}{{Munshi} et~al.}{2024}]{Munshi_nenuFAR_2024}
{Munshi} S.,  et~al., 2024, \mn@doi [\aap] {10.1051/0004-6361/202348329}, \href {https://ui.adsabs.harvard.edu/abs/2024A&A...681A..62M} {681, A62}

\bibitem[\protect\citeauthoryear{Neben et~al.,}{Neben et~al.}{2015}]{neben_measuring_2015}
Neben A.~R.,  et~al., 2015, \mn@doi [Radio Science] {10.1002/2015RS005678}, 50, 2015RS005678

\bibitem[\protect\citeauthoryear{Neben et~al.,}{Neben et~al.}{2016}]{neben_hydrogen_2016}
Neben A.~R.,  et~al., 2016, arXiv:1602.03887 [astro-ph]

\bibitem[\protect\citeauthoryear{Ninni, Bolli, Paonessa, Pupillo, Virone  \& Wijnholds}{Ninni et~al.}{2020a}]{ninni_comparison_2020}
Ninni P.~D.,  Bolli P.,  Paonessa F.,  Pupillo G.,  Virone G.,   Wijnholds S.~J.,  2020a, in 2020 14th {European} {Conference} on {Antennas} and {Propagation} ({EuCAP}). pp~1--5, \mn@doi{10.23919/EuCAP48036.2020.9135792}, \url {https://ieeexplore.ieee.org/document/9135792}

\bibitem[\protect\citeauthoryear{{Ninni}, {Bolli}, {Paonessa}, {Pupillo}, {Virone}  \& {Wijnholds}}{{Ninni} et~al.}{2020b}]{Ninni_drones_lofar_2020}
{Ninni} P.~D.,  {Bolli} P.,  {Paonessa} F.,  {Pupillo} G.,  {Virone} G.,   {Wijnholds} S.~J.,  2020b, in 2020 14th European Conference on Antennas and Propagation (EuCAP). pp~1--5, \mn@doi{10.23919/EuCAP48036.2020.9135792}

\bibitem[\protect\citeauthoryear{Nunhokee et~al.,}{Nunhokee et~al.}{2020}]{nunhokee_measuring_2020}
Nunhokee C.~D.,  et~al., 2020, \mn@doi [The Astrophysical Journal] {10.3847/1538-4357/ab9634}, 897, 5

\bibitem[\protect\citeauthoryear{{Oh} \& {Mack}}{{Oh} \& {Mack}}{2003}]{oh_peng_foregrounds_2003}
{Oh} S.~P.,  {Mack} K.~J.,  2003, \mn@doi [\mnras] {10.1111/j.1365-2966.2003.07133.x}, \href {https://ui.adsabs.harvard.edu/abs/2003MNRAS.346..871O} {346, 871}

\bibitem[\protect\citeauthoryear{Orosz, Dillon, Ewall-Wice, Parsons  \& Thyagarajan}{Orosz et~al.}{2019}]{orosz_mitigating_2019}
Orosz N.,  Dillon J.~S.,  Ewall-Wice A.,  Parsons A.~R.,   Thyagarajan N.,  2019, \mn@doi [Monthly Notices of the Royal Astronomical Society] {10.1093/mnras/stz1287}, 487, 537

\bibitem[\protect\citeauthoryear{{Paonessa}, {Ciorba}, {Virone}, {Bolli}, {Magro}, {McPhail}, {Minchin}  \& {Bhushan}}{{Paonessa} et~al.}{2020}]{Paonessa_drones_ska_2020}
{Paonessa} F.,  {Ciorba} L.,  {Virone} G.,  {Bolli} P.,  {Magro} A.,  {McPhail} A.,  {Minchin} D.,   {Bhushan} R.,  2020, in 2020 XXXIIIrd General Assembly and Scientific Symposium of the International Union of Radio Science. pp~1--3, \mn@doi{10.23919/URSIGASS49373.2020.9232190}

\bibitem[\protect\citeauthoryear{{Parsons}, {Pober}, {Aguirre}, {Carilli}, {Jacobs}  \& {Moore}}{{Parsons} et~al.}{2012}]{parsons_wedge_2012}
{Parsons} A.~R.,  {Pober} J.~C.,  {Aguirre} J.~E.,  {Carilli} C.~L.,  {Jacobs} D.~C.,   {Moore} D.~F.,  2012, \mn@doi [\apj] {10.1088/0004-637X/756/2/165}, \href {https://ui.adsabs.harvard.edu/abs/2012ApJ...756..165P} {756, 165}

\bibitem[\protect\citeauthoryear{Patil et~al.,}{Patil et~al.}{2016}]{patil_systematic_2016}
Patil A.~H.,  et~al., 2016, \mn@doi [Monthly Notices of the Royal Astronomical Society] {10.1093/mnras/stw2277}, 463, 4317

\bibitem[\protect\citeauthoryear{{Pober} et~al.,}{{Pober} et~al.}{2013}]{pober_wedge_2013}
{Pober} J.~C.,  et~al., 2013, \mn@doi [\apjl] {10.1088/2041-8205/768/2/L36}, \href {https://ui.adsabs.harvard.edu/abs/2013ApJ...768L..36P} {768, L36}

\bibitem[\protect\citeauthoryear{{Pober} et~al.,}{{Pober} et~al.}{2014}]{pober_next_gen_eor_2014}
{Pober} J.~C.,  et~al., 2014, \mn@doi [\apj] {10.1088/0004-637X/782/2/66}, \href {https://ui.adsabs.harvard.edu/abs/2014ApJ...782...66P} {782, 66}

\bibitem[\protect\citeauthoryear{{Pober} et~al.,}{{Pober} et~al.}{2016}]{pober_foreground_removal_2016}
{Pober} J.~C.,  et~al., 2016, \mn@doi [\apj] {10.3847/0004-637X/819/1/8}, \href {https://ui.adsabs.harvard.edu/abs/2016ApJ...819....8P} {819, 8}

\bibitem[\protect\citeauthoryear{Rahimi et~al.,}{Rahimi et~al.}{2021}]{rahimi_epoch_2021}
Rahimi M.,  et~al., 2021, \mn@doi [Monthly Notices of the Royal Astronomical Society] {10.1093/mnras/stab2918}, 508, 5954

\bibitem[\protect\citeauthoryear{{Salvini} \& {Wijnholds}}{{Salvini} \& {Wijnholds}}{2014}]{Salvini_Wijnholds_Cal_2014}
{Salvini} S.,  {Wijnholds} S.~J.,  2014, \mn@doi [\aap] {10.1051/0004-6361/201424487}, \href {https://ui.adsabs.harvard.edu/abs/2014A&A...571A..97S} {571, A97}

\bibitem[\protect\citeauthoryear{{Santos}, {Cooray}  \& {Knox}}{{Santos} et~al.}{2005}]{santos_eor_2005}
{Santos} M.~G.,  {Cooray} A.,   {Knox} L.,  2005, \mn@doi [\apj] {10.1086/429857}, \href {https://ui.adsabs.harvard.edu/abs/2005ApJ...625..575S} {625, 575}

\bibitem[\protect\citeauthoryear{{Sokolowski} et~al.,}{{Sokolowski} et~al.}{2017}]{Sokolowski_FEE_2017}
{Sokolowski} M.,  et~al., 2017, \mn@doi [\pasa] {10.1017/pasa.2017.54}, \href {https://ui.adsabs.harvard.edu/abs/2017PASA...34...62S} {34, e062}

\bibitem[\protect\citeauthoryear{Sutinjo, O'Sullivan, Lenc, Wayth, Padhi, Hall  \& Tingay}{Sutinjo et~al.}{2015}]{Sutinjo_FEE_2015}
Sutinjo A.~T.,  O'Sullivan J.,  Lenc E.,  Wayth R.~B.,  Padhi S.,  Hall P.~A.,   Tingay S.~J.,  2015, Radio Science, 50, 52

\bibitem[\protect\citeauthoryear{{Thyagarajan} et~al.,}{{Thyagarajan} et~al.}{2013}]{thyagarajan_wedge_2013}
{Thyagarajan} N.,  et~al., 2013, \mn@doi [\apj] {10.1088/0004-637X/776/1/6}, \href {https://ui.adsabs.harvard.edu/abs/2013ApJ...776....6T} {776, 6}

\bibitem[\protect\citeauthoryear{{Thyagarajan} et~al.,}{{Thyagarajan} et~al.}{2015}]{thyagarajan_wedge_2015}
{Thyagarajan} N.,  et~al., 2015, \mn@doi [\apj] {10.1088/0004-637X/804/1/14}, \href {https://ui.adsabs.harvard.edu/abs/2015ApJ...804...14T} {804, 14}

\bibitem[\protect\citeauthoryear{Tingay et~al.,}{Tingay et~al.}{2013}]{Tingay_MWA_2013}
Tingay S.~J.,  et~al., 2013, \mn@doi [Publications of the Astronomical Society of Australia] {10.1017/pasa.2012.007}, 30, e007

\bibitem[\protect\citeauthoryear{{Trott}, {Wayth}  \& {Tingay}}{{Trott} et~al.}{2012}]{trott_wedge_2012}
{Trott} C.~M.,  {Wayth} R.~B.,   {Tingay} S.~J.,  2012, \mn@doi [\apj] {10.1088/0004-637X/757/1/101}, \href {https://ui.adsabs.harvard.edu/abs/2012ApJ...757..101T} {757, 101}

\bibitem[\protect\citeauthoryear{{Trott} et~al.,}{{Trott} et~al.}{2016}]{Trott_CHIPS_2016}
{Trott} C.~M.,  et~al., 2016, \mn@doi [\apj] {10.3847/0004-637X/818/2/139}, \href {https://ui.adsabs.harvard.edu/abs/2016ApJ...818..139T} {818, 139}

\bibitem[\protect\citeauthoryear{Trott et~al.,}{Trott et~al.}{2020}]{trott_deep_2020}
Trott C.~M.,  et~al., 2020, \mn@doi [Monthly Notices of the Royal Astronomical Society] {10.1093/mnras/staa414}, 493, 4711

\bibitem[\protect\citeauthoryear{{Vedantham}, {Udaya Shankar}  \& {Subrahmanyan}}{{Vedantham} et~al.}{2012}]{vedantham_wedge_2012}
{Vedantham} H.,  {Udaya Shankar} N.,   {Subrahmanyan} R.,  2012, \mn@doi [\apj] {10.1088/0004-637X/745/2/176}, \href {https://ui.adsabs.harvard.edu/abs/2012ApJ...745..176V} {745, 176}

\bibitem[\protect\citeauthoryear{Wayth et~al.,}{Wayth et~al.}{2018}]{Wayth_MWA_2018}
Wayth R.~B.,  et~al., 2018, \mn@doi [Publications of the Astronomical Society of Australia] {10.1017/pasa.2018.37}, 35, e033

\bibitem[\protect\citeauthoryear{{Wilensky} et~al.,}{{Wilensky} et~al.}{2024}]{wilensky_beam_2024}
{Wilensky} M.~J.,  et~al., 2024, \mn@doi [arXiv e-prints] {10.48550/arXiv.2403.13769}, \href {https://ui.adsabs.harvard.edu/abs/2024arXiv240313769W} {p. arXiv:2403.13769}

\bibitem[\protect\citeauthoryear{{Yatawatta} et~al.,}{{Yatawatta} et~al.}{2013}]{yatawatta_lofar_eor_2013}
{Yatawatta} S.,  et~al., 2013, \mn@doi [\aap] {10.1051/0004-6361/201220874}, \href {https://ui.adsabs.harvard.edu/abs/2013A&A...550A.136Y} {550, A136}

\bibitem[\protect\citeauthoryear{Zarka et~al.,}{Zarka et~al.}{2020}]{zarka_low-frequency_2020}
Zarka P.,  et~al., 2020, in 2020 URSI General Assembly and Scientific Symposium.

\makeatother
\end{thebibliography}








\bsp	
\label{lastpage}
\end{document}